          [1999/12/02 1.01 (PWD)]
\documentclass[letterpaper,twocolumn]{esapub} % American paper

\usepackage{natbib,graphicx}

\title{Annihilation Radiation in the Galaxy}
\author{C. D. Dermer}
\author{R. J. Murphy}
\affil{Code 7650, Naval Research Laboratory, Washington, DC 20375-5352, USA}

\newcommand{\psim}{\lower.5ex\hbox{$\; \buildrel \propto \over\sim \;$}}
\def\gtrsim{\lower.5ex\hbox{$\; \buildrel > \over\sim \;$}} 
\def\lesssim{\lower.5ex\hbox{$\; \buildrel < \over\sim \;$}} 

\begin{document}


\maketitle

\begin{abstract}
Observations of annihilation radiation in the Galaxy are briefly
reviewed. We summarize astrophysical mechanisms leading to positron
production and recent estimates for production rates from nova and
supernova nucleosynthesis in the Galaxy. The physical processes
involved in the production of annihilation radiation in the
interstellar medium are described.  These include positron
thermalization, charge exchange, radiative recombination, and direct
annihilation.  Calculations of 2$\gamma$ and 3$\gamma$ spectra and the
positronium (Ps) fraction due to the annihilation of positrons in
media containing H and He at different temperatures and ionization
states are presented. Quenching of Ps by high temperature plasmas or
dust could account for differences between 0.511 MeV and $3\gamma$ Ps
continuum maps. These results are presented in the context of the
potential of INTEGRAL to map sites of annihilation radiation in the
Galaxy.  Positron production by compact objects is also considered.
\end{abstract}

\section{Introduction}

With an omnidirectional flux of $\approx 2.6 \times 10^{-3}$ ph
cm$^{-2}$ s$^{-1}$, the e$^+$-$e^-$ annihilation line at 0.511 MeV is
the brightest gamma-ray line in the Galaxy. Associated with the
$2\gamma$ line is the $3\gamma$ positronium continuum formed by the
decay of Positronium (Ps) atoms in the $^3$S$_1$ state. The fraction
of positrons that annihilate via the formation of Ps depends on the
temperature, density, composition, and dust content of the medium.
Consequently observations of $2\gamma$ and $3\gamma$ fluxes and line
widths reveal the ISM properties at sites where positrons
annihilate. Past missions have provided limited information on ISM
conditions at localized sites because they employed non-imaging
detectors with large fields-of-view (FoVs). Moreover, many earlier
annihilation line observations used scintillation detectors rather
than germanium spectrometers, and so were not able to resolve the
width of the 0.511 MeV line. The capability of {\it INTEGRAL} to
resolve and image the annihilation line --- though not without
considerable technical hurdles --- promises to revolutionize our view
of the Galaxy in the light of the e$^+$-$e^-$ annihilation line.

This review summarizes the status of annihilation line astrophysics in
the period following the demise of the {\it Compton Gamma Ray
Observatory} and preceding the launch of {\it INTEGRAL}. In Section 2,
past observations of annihilation radiation in the Galaxy are briefly
reviewed. Section 3 summarizes the principal astrophysical processes
that yield positrons, including recent calculations of positron
production by novae, supernovae, cosmic rays, and galactic compact
objects. Physical processes leading to the 2$\gamma$ line and
3$\gamma$ continuum are summarized in Section 4, and new calculations
of line and continuum spectra and 3$\gamma$/$2\gamma$ flux ratios in
different media are presented in Section 5.  The review is concluded
in Section 6 with a discussion of prospects for {\it INTEGRAL} to
identify the origin of positrons that produce annihilation radiation,
including positron production by Galactic compact objects.

\section{Spectra of Galactic Annihilation Radiation}

The spectrum of the Galaxy as measured by a soft $\gamma$-ray
telescope is complicated not only by the internal and particle
backgrounds of the detector, but also by variable Galactic
$\gamma$-ray point sources and the unknown distribution of diffuse
emissions in the Galaxy. These technical difficulties have been dealt
with in a variety of ways, leading to our present incomplete picture
of the annihilation emission of the Galaxy.

The 0.511 MeV line was originally discovered by the Rice University
group \citep{jh73}, though positive identification of the line
required the flight of a germanium detector \citep{lev78} which showed
that the line, averaged over the 15$^\circ$ FoV of the Bell-Sandia
telescope, has a FWHM width $\Delta E < 3.2$ keV. Other observations
of annihilation radiation have employed broad field-of-view
instruments, the largest being the 130$^\circ$ FWHM FoV of the {\it
Solar Maximum Mission} Gamma Ray Spectrometer \citep{sha88}. By
showing that the measured flux correlates with the opening angle of
the detector, \citet{lr89} firmly established that there exists a
diffuse component of the annihilation radiation in the Galaxy. They
also argued for the existence of a variable, time-dependent source of
annihilation radiation on the basis of multiple observations made with
the HEAO-3 and Bell-Sandia germanium spectrometers. Later analysis of
the HEAO-3 data \citep{mlw94} showed that the evidence for a variable
point source was marginal, and no evidence for a variable annihilation
component has been obtained through 5.5 \citep{pur97} and 8.5
\citep{mil99} years of observations of the Galactic Center region with
the OSSE instrument on {\it Compton GRO}.

OSSE has been used to measure the Galactic annihilation radiation in
the greatest detail (Milne et~al., these proceedings). With a FWHM FoV
of 11.4$^\circ\times 3.8^\circ$, representing about 0.1\% of the full
sky, OSSE can map annihilation radiation through multiple observations
in different directions. This requires offset pointing to subtract
background and sophisticated mapping techniques to map diffuse
emission. An observation lasting for two 2-week periods is shown in
Fig.\ 1, and illustrates the dominant features in the spectrum of a
Galactic Center pointing. Shown is the spectrum measured during
Viewing Periods 5 and 16 \citep{kin99}. The prominent 0.511 MeV
annihilation line has a flux of $\phi_{2\gamma} = 1.8 \pm 0.19 \times
10^{-4}$ ph cm$^{-2}$ s$^{-1}$.  The deduced flux of the $3\gamma$
continuum is $\phi_{3\gamma} = 10.50 \pm 0.64,pur97 \times 10^{-4}$ ph
cm$^{-2}$ s$^{-1}$.

\begin{figure}
\centering
\rotatebox{90}{\includegraphics[width=0.75\linewidth]{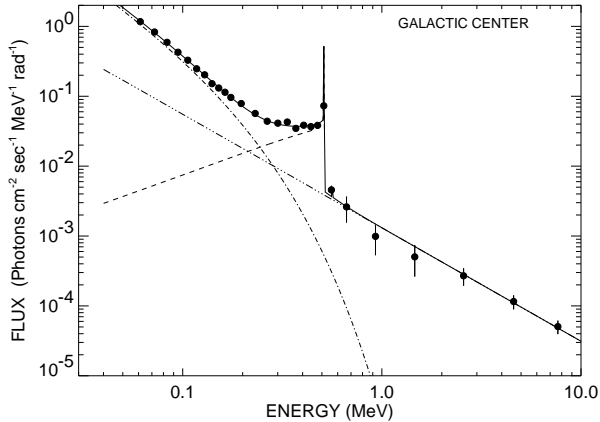}}
\caption{OSSE spectrum of the Galactic Center region 
during Viewing Periods 5 and 16,
showing the 0.511 MeV line and the residual radiation separated into
the 3$\gamma$ positronium continuum and background emissions.
\label{fig:spectrum}}
\end{figure}

The omnidirectional flux of 0.511 MeV annihilation radiation measured
with the {\it Solar Maximum Mission} is $\phi_{2\gamma} =
23^{+5}_{-8}\times 10^{-4}$ ph cm$^{-2}$ s$^{-1}$, though the precise
flux measurement depends on whether the underlying distribution of the
annihilation radiation is assumed to be described by a Galactic Center
point source or a distribution described by, for example, the galactic
CO emission \citep{sha90,har90}. The omnidirectional flux inferred
\citep{pur97} through analyses of OSSE and earlier annihilation line
observations is $\phi_{2\gamma} = 28\pm 4\times 10^{-4}$ ph cm$^{-2}$
s$^{-1}$, consistent with the {\it SMM} result.

Naively assuming that the bulk of the annihilation takes place at the
distance of the Galactic Center implies a total Galactic annihilation
rate of $\dot N_{2\gamma} \approx 4\pi (8$ kpc$)^2 \phi_{2\gamma}
\approx 2\times 10^{43}$ ph s$^{-1}$. Letting $f$ represent the
fraction of positrons that annihilate via the intermediate formation
of Ps, then each $e^+$ produced $2[(1-f)+ (f/4)]$ annihilation
photons, because a Ps atom decays 1/4 of the time via the $2\gamma$
state.  If $f \cong 0.9$, as implied by spectral analyses, then
positron sources in the Milky Way are producing a time-averaged rate
$\approx 3\times 10^{43}$ e$^+$ s$^{-1}$.

The structure in the distribution of annihilation radiation can be
determined either through model-dependent \citep{ski93} or
model-independent \citep{pur97,che97,mil99} mapping techniques. In the
former approach, one assumes that the annihilation radiation is
described by model spatial distributions, for example, those
describing novae, pulsars, hot plasma, or $> 100$ MeV gamma-ray
distributions. One then minimizes $\chi^2$ of the observations with
respect to various superpositions of the model distributions. This
method is limited if the model distributions provide a poor
characterization of features peculiar to the true spatial distribution
of annihilation radiation. These methods were used to argue that a
disk component, a bulge feature related to an old stellar population,
and a time-variable point source were required to fit available 0.511
MeV line observations \citep{rsl94}.

Model-independent mapping techniques, employing Singular Value
Decomposition or the Maximum Entropy Method algorithms, can be used to
reconstruct the image without recourse to assumptions about underlying
model distributions. Using these methods, \citet{pur97} showed that
OSSE observations were described by the following emission components:
\begin{itemize}
\item Positive galactic plane: $2.9\pm 0.2$,
\item Negative galactic plane: $3.1\pm 0.3$,
\item Central bulge: $6.0\pm 0.4$,
\item Positive latitude enhancement: $2.2\pm 0.2$,
\item Mirror region: $0.8\pm 0.1$ ,
\end{itemize}
where the fluxes are given in units of $10^{-4}$ ph cm$^{-2}$
s$^{-1}$. This analysis results in a flux for the full map of $22.5\pm
0.7$ ph cm$^{-2}$ s$^{-1}$, which is consistent with the
omnidirectional flux quoted earlier when consideration of emission in
regions not observed with OSSE was included. The asymmetry associated
with the positive latitude enhancement has been interpreted
\citep{ds97} in terms of annihilation fountain of pair-laden hot
plasma driven by a starburst from the Galactic Center region into the
halo of the Galaxy.

Later analysis \citep{mil99} has confirmed that the positive latitude
enhancement has a flux $\phi_\gamma \cong 1$-2$\times 10^{-4}$ ph
cm$^{-2}$ s$^{-1}$, though there is only marginal evidence for a
high-latitude enhancement in the $3\gamma$ positronium map. This
implies that Ps is quenched in the fountain, either because the
temperature is very high or the region is dusty. The first possibility
can be determined by measuring the line width of the annihilation
radiation in this direction, and will be discussed in more detail in
the Section 5.

\section{Positron Processes and Production Sites}

\begin{table}
  \begin{center}
    \caption{Positron production processes and mean production 
Lorentz factors}\vspace{1em}
    \renewcommand{\arraystretch}{1.2}
    \begin{tabular}[h]{lc}
      \hline
      Processes &   $\langle \gamma \rangle $    \\
      \hline
      N$^* \rightarrow$ N+e$^+$  &  few \\
      N + p $ \rightarrow \pi^+ \rightarrow$ e$^+$   & $\gtrsim 60$  \\
      N + N$^\prime \rightarrow$ N$^* \rightarrow$ N + e$^+$  & 
$\gtrsim$ few  \\
      $e+B \rightarrow \gamma + B \rightarrow$ e$^+$e$^-$   &
 $\sim 10^6$-$10^8$  \\
      $\gamma\gamma \rightarrow$ e$^+$e$^-$  & $\gtrsim$ few  \\
      \hline \\
      \end{tabular}
    \label{tab:table}
  \end{center}
\end{table}

As we have seen, sources must produce $\approx 3\times 10^{43}$ e$^+$
s$^{-1}$ to account for the diffuse annihilation glow of the Milky
Way. Table 1 lists astrophysical processes that make positrons. The
first row represents channels where e$^+$ are produced through
$\beta^+$ decay of nuclei formed in explosive and hydrostatic
nucleosynthesis of novae, supernovae (SNe), Wolf-Rayet and Asymptotic
Giant Branch stars. The positrons produced through this channel have a
characteristic $\beta^+$ decay spectrum with typical kinetic energies
of $\sim 1 $ MeV. The second row represents secondary production
interactions between cosmic ray protons and ions that produce a
characteristic $\pi^+$ decay spectrum of positrons peaking at $\sim$
30 MeV. Positrons can also be produced in stellar flares through this
process.  The third row represents cosmic ray interactions that form
ions which subsequently decay through the emission of
positrons. Depending on whether the ions are at rest or are in motion,
positrons are formed with Lorentz factors ranging from only a few to
values roughly equal to the Lorentz factors of the excited ion. This
process can be important in stellar flares and in accreting neutron
stars and black holes, and in regions with enhanced fluxes of
low-energy cosmic rays.

The fourth row represents positrons formed by pulsar electromagnetic
cascades. Here curvature or synchrotron photons materialize in the
strong magnetic field near a pulsar polar cap. These positrons can
escape through magnetic fields that open through the light
cylinder. The Lorentz factors of positrons made in this way are,
however, usually so large that they will not thermalize on a time
scale corresponding to the age of the Galaxy. The $\gamma\gamma$ pair
production process of the fifth row occurs in luminous compact regions
that may be found in galactic black hole candidates, microquasars,
active galactic nuclei, and in gamma-ray bursts (Dermer \& B\"ottcher,
these proceedings).  Not listed in Table 1 are higher-order processes
such as triplet pair production $\gamma +$ e $\rightarrow $ e+e$^+$
+e$^-$, and direct pair production e+ N $\rightarrow $ e+ N + e$^+$ +
e$^-$, which are generally much less important because of their small
cross sections.

\subsection{Positron Production in Novae}
There are four principal channels for the production of positrons in
Novae; these are,
\begin{enumerate}
\item $^{13}$Ni $\rightarrow ^{13}$ C, $\tau = 598$ s, 100\% $\beta^+$
\item $^{18}$F $\rightarrow ^{18}$ O, $\tau = 109.8$ min, 96.9\% $\beta^+$
\item $^{22}$Na $\rightarrow ^{22}$ Ne, $\tau = 2.60$ yr, 90.4\% $\beta^+$
\item $^{26}$Al $\rightarrow ^{26}$ Mg, $\tau = 7.2\times 10^5$ yr, 82\% $\beta^+$ and $\gamma$(1.809 MeV),
\end{enumerate}
where we also list the half-life $\tau$ and the fraction of decays
that yield positrons. Nucleosynthesis in novae is discussed at this
conference by Hernanz; here we only note that the first two channels
of $\beta^+$ production are sufficiently prompt that the produced
positrons will annihilate in the dense material formed in the nova
explosion \cite{her99}, whereas e$^+$ made through the latter two
channels will be mixed in the expanding envelope and can be injected
into the ISM. The Galaxy-averaged production rate of novae from
Oxygen/Neon white dwarfs is estimated by \citep{her99a} to be $\dot
N_+^{22} \cong 5\times 10^{39}$ - $2.4\times 10^{40} \dot N_{ONe/yr}$
e$^+$ s$^{-1}$ and $\dot N_+^{26} \cong 7\times 10^{38}$ - $4\times
10^{40} \dot N_{ONe/yr}$ e$^+$ s$^{-1}$ for the $^{22}$Na and
$^{26}$Al channels, respectively. The quantity $\dot N_{ONe/yr}$ is
the yearly rate of novae throughout the Galaxy. \citet{sta97} give
$\dot N_+^{22} \cong 8\times 10^{38}$ - $5\times 10^{40} \dot
N_{ONe/yr}$ e$^+$ s$^{-1}$ and $\dot N_+^{26} \cong 6\times 10^{39}$ -
$2\times 10^{41} \dot N_{ONe/yr}$ e$^+$ s$^{-1}$ for these rates,
which are slightly larger but consistent given the many uncertainties
in the calculation.  They also calculate the e$^+$ production rate
from novae on CO white dwarfs, and obtain $\dot N_+^{22} \cong 6\times
10^{36}$ - $6\times 10^{37} \dot N_{CO/yr}$ e$^+$ s$^{-1}$ and $\dot
N_+^{26} \cong 7\times 10^{36}$ - $8\times 10^{37} \dot N_{CO/yr}$
e$^+$ s$^{-1}$. The yearly rate of novae throughout the Galaxy is
estimated from observations to be $\approx 35$ per year. If a
significant fraction of these novae are produced on O/Ne white dwarfs,
then novae can provide a nonnegligible amount of the positrons
injected into the Galaxy and would contribute to both the disk and
bulge annihilation radiation.

\subsection{Positron Production from SNe and Stellar Winds}

The dominant channels for positron production and the associated
$\gamma$ ray lines formed through hydrostatic and explosive
nucleosynthesis are
\begin{enumerate}
\item $^{57}$Ni$\rightarrow ^{57}$Co, $\tau = 36$ hr, 40\% $\beta^+$
 and $\gamma$(1370 keV), followed by
$^{57}$Co $\rightarrow ^{57}$Fe, $\tau = 271$ d, $\gamma$(122, 136 keV);
\item $^{56}$Ni$\rightarrow ^{56}$Co, $\tau = 6.1$ d, $\gamma$(150, 750, 
812 keV), followed by $^{56}$Co $\rightarrow ^{56}$Fe, $\tau =$ 78.8 d,
 18\% $\beta^+$ $\gamma$(847, 1238 keV);
\item $^{44}$Ti$\rightarrow ^{44}$Sc, $\tau \cong 62$ yr, $\gamma$(67.8, 
78.4 keV), followed by $^{44}$Sc$\rightarrow ^{44}$Ca, $\tau = 3.93$ yr,
 95\% $\beta^+$ and $\gamma$(1.16 MeV);
\item $^{26}$Al$\rightarrow ^{26}$ Mg, $\tau = 7.2\times 10^5$ yr,
 82\% $\beta^+$ and $\gamma$(1.809 MeV).
\end{enumerate}
The $\beta^+$ decay lifetime from the channel 1 is too short to permit
the positrons to escape from the ejecta into the ISM, so only a prompt
0.511 MeV line could be formed, and this would be severely degraded in
optically thick material. The positron production rates into the ISM
from Type Ia and Type II supernovae for the remaining channels depend
sensitively on the percentage $\eta_{-2}$ of positrons that escape
into the ISM.  For channel 2, \citet{cl93} find that $\dot N_+^{56,Ia}
\cong 1.4\times 10^{43} \eta_{-2} M_{56} \dot N_{Ia/C}$ e$^+$
s$^{-1}$, where $M_{56}$ is the average number of solar masses of
synthesized $^{56}$Fe per SN, and $\dot N_{Ia/C}$ is the mean number
of Type Ia SNe per century in the Milky Way. \citet{ww92} calculate
that $0.6 < M_{56} < 0.9$. From these rates, it is clear that Type Ia
SNe are capable of making an important if not dominant contribution to
positron production, depending crucially on the escape fraction
$\eta_{-2}$ \citep{mtl99}.

Positron production through the $^{56}$Ni$\rightarrow
^{56}$Co$\rightarrow ^{56}$Fe channel in Type II SNe is estimated by
\citet{cl93} to be $\dot N_+^{56,II} \approx \times 10^{42} \eta_{-2}
(M_{56}/0.08) \dot N_{II/C}$ e$^+$ s$^{-1}$. Because of the thickness
of the overlying envelope in Type II SNe, the escape fraction is very
poorly known and is probably $\ll 1$\%. At best, this channel could
make a minor contribution to the total positron production rate in the
Galaxy.

The much smaller synthesized mass of $^{44}$Ti compared with $^{56}$Fe
means that the $^{44}$Ti$\rightarrow ^{44}$Sc$\rightarrow ^{44}$Ca
channel makes a small contribution to the Galactic positron production
rate from Type Ia SNe, even taking into account the longer decay time
scale and larger positron escape fraction. The production rate from
Type Ib and II SNe depends on the large uncertainties in the amount of
synthesized mass per SNe. In the extreme case, the positron production
rate from Type Ia, Ib, and II SNe is estimated to be $\dot N_+^{44}
\simeq 10^{43}$ e$^+$ s$^{-1}$ \citep{cl93}.

All positrons produced through the $^{26}$Al$\rightarrow ^{26}$Mg
chain escape into the ISM due to the long lifetime of $^{26}$Al. The
1.809 MeV line therefore provides a tracer of positron production
through channel 4. With an omnidirectional 1.809 MeV flux of $\approx
3\times 10^{-4}$ cm$^{-2}$ s$^{-1}$ \citep{die95}, this implies an
associated Galactic positron production rate of $\sim 2\times 10^{42}$
e$^+$ s$^{-1}$.  Type II SNe produce positrons from $^{26}$Al at a
rate of $\dot N_+^{26} \approx 10^{41} (M_{26}/10^{-5}) \dot N_{II/C}$
e$^+$ s$^{-1}$, where the mass of $^{26}$Al produced per SN ranges
from $0.3 \lesssim M_{26}/10^{-5} \lesssim 20$ \citep{pra96}. It is
therefore likely that other sources contribute to $^{26}$Al
production, such as massive stellar winds and novae. Winds from
Wolf-Rayet stars produce positrons at the rate of $\dot N_+^{26}
\approx 1\times 10^{42}$ e$^+$ s$^{-1}$ \citep{mey97}. AGB stars
(Mowlavi, these proceedings) and novae can also make significant
contributions. Thus the annihilation line map of the Galaxy should
display a component that traces the 1.809 MeV map, although transport
of the positrons from the production to the annihilation site could
modify the spatial correlation, as would be the case for the positive
latitude enhancement if due to a starburst/superwind \citep{ds97}.

\subsection{Other Positron Sources}

Cosmic rays produce positrons through the same strong interactions
that produce $\pi^0 $ gamma rays. Thus the $\pi^0$ $\gamma$-ray
emission centered at $\approx 70$ MeV provides a tracer of positron
production. COS-B analyses \citep{blo84} imply that the Galaxy
produces $\sim 2.5\times 10^{42}$ $>$100 MeV ph s$^{-1}$. The $\pi^0$
and $\pi^+$ production cross sections are roughly equal, though two
$\gamma$-ray photons are made from each $\pi^0$. Thus the $>$100 MeV
Galactic gamma-ray production rate also roughly represents an upper
limit to the positron production rate from cosmic ray interactions,
after correcting for the emission below 100 MeV and noting that
bremsstrahlung and Compton photons substantially contribute to the
diffuse Galactic gamma-ray background. Large fluxes of low-energy
cosmic rays would also contribute to positron production through
$\beta^+$ decay of radioactive nuclei. COMPTEL upper limits to nuclear
deexcitation gamma rays \citep{blo99} suggest that positrons made
through this process do not, however, significantly contribute to the
overall Galactic annihilation line flux.

Although difficult to quantify, compact objects could eject large
numbers of positrons into the ISM. Galactic black-hole candidate
sources \citep{dl88} and microquasars \citep{ll96} are the two leading
candidates for producing localized and time-variable sites of
annihilation radiation. Individual sources could eject $\sim
10^{41}$-$10^{42}$ e$^+$ s$^{-1}$ into the ISM. Evidence has been
presented from Sigma/Granat observations of 1E 1740.7-2942 and Nova
Muscae for episodes of line emission at $\approx 400$ keV that was
attributed to redshifted annihilation radiation \citep{har97}, but
OSSE has not detected variable annihilation radiation or annihilation
lines from black-hole candidate sources such as Cygnus X-1
\citep{phl96}. As already noted, electromagnetic cascades in pulsars
would probably produce positrons with Lorentz factors too great to
thermalize within the lifetime of the Galaxy. Production of positrons
by GRBs in the Galaxy is discussed in a separate contribution (Dermer
\& B\"ottcher, these proceedings).

\section{Positron Annihilation Processes}

Early calculations of positron annihilation in the ISM were treated by
\citet{bus79}, and in Solar flares by \citet{cra76}. Updated cross
sections and processes related to the annihilation of positrons are
discussed in detail elsewhere \citep{grl91,gsr97}, so they are
reviewed only briefly here. Positrons injected into the ISM will
annihilate directly or thermalize. Thermalization is more important
than annihilation for positrons injected at MeV energies, and direct
annihilation of higher energy positrons produces a line that is
invariably too broadened to be detectable. Positrons thermalize
through Coulomb losses with free electrons, and through excitation and
ionization energy losses with bound electrons. When a positron's
energy reaches a few hundred eV, the energy-loss processes compete
primarily with the formation of Ps through charge exchange in flight,
as the rate of direct annihilation-in-flight is small.  Thermal
positrons either annihilate directly with free or bound electrons, or
form Ps through radiative recombination and charge exchange. Charge
exchange with neutral hydrogen has a 2430 \AA ~threshold, so this
process is suppressed in sufficiently low-temperature gases.

Ps is formed in the triplet ($^3S_1$) ortho-Ps and singlet ($^1S_0$)
para-Ps ground states in a 3:1 ratio, with decay lifetimes of
$1.4\times 10^{-7}$ s and $1.25\times 10^{-10}$ s,
respectively. Cascading to the ground state can produce a Ly-$\alpha$
line at 6.8 eV and other lines that are very weak but, in principle,
detectable \citep{bur94,wal96}. Ps can be quenched if the density is
$\gtrsim 10^{13}$ cm$^{-3}$, though this is not relevant for
annihilation in the ISM. Annihilation of thermalized positrons on dust
\citep{zur85,grl91} is also an important quenching process for Ps,
leading to a suppression of the $3\gamma$ continuum. These
calculations are complicated by grain properties, including their
sizes, charges, and compositions. \citet{grl91} find that the presence
of dust in the warm envelopes around cold molecular cloud cores can
reduce the Ps fraction, but that dust is unlikely to affect the
annihilation line and continuum properties in the uniform ISM.

\section{Calculations of Annihilation Line and Continuum Spectra}

We have developed a computer code to calculate the 0.511 MeV line
spectrum and relative strength of the $3\gamma$ continuum. The code
incorporates updated cross sections to calculate thermally-averaged
rates for all relevant binary interactions involving positrons, and
uses a Monte Carlo technique to simulate the thermalization and Ps
production of $\sim$MeV positrons. We include charge exchange with
neutral and singly-ionized helium, which have generally been ignored
in astrophysical applications. Processes involving dust and $H_2$ are
not yet included.

\begin{table}[t]
  \begin{center}
    \caption{Binary interactions involving positrons.}\vspace{1em}
    \renewcommand{\arraystretch}{1.2}
    \begin{tabular}[h]{ll}
      \hline
      Energy Redistribution &   Positronium Formation  \\
      \hline
$e^+$ H $\rightarrow e^+$ H$^*$ &
$e^+\,e^-\rightarrow \gamma$ Ps \\
$e^+$ H $\rightarrow e^+e^-$ H$^+$ &
$e^+\,$ H $\rightarrow$ H$^+$ Ps \\
$e^+$ He $\rightarrow e^+$ He$^*$  &
$e^+\,$ He $\rightarrow$ He$^+$ Ps  \\
$e^+$ He $\rightarrow e^+e^-$ He$^+$  & \\
      \hline \\
      \hline
       Positronium Quenching &  Annihilation \\
      \hline
$e^-\;^3$Ps $\rightarrow e^-e^-e^+$ &
$e^+\,e^-\rightarrow 2\gamma$ \\
H $^3$Ps $\rightarrow$ H $e^-e^+$ &
$e^+$ H $\rightarrow$ H$^+$ $2\gamma$ \\
$e^-\;^3$Ps $\rightarrow e^-\;^1$Ps &
$e^+$ He $\rightarrow$ He$^+$ $2\gamma$ \\
H $^3$Ps $\rightarrow$ H $^1$Ps &
$^1$Ps $\rightarrow 2\gamma$ \\
H$^+$ Ps $\rightarrow$ H $e^+$& $^3$Ps $\rightarrow 3\gamma$ \\
He$^+$ Ps $\rightarrow$ He $e^+$ &\\
      \hline \\
      \end{tabular}
    \label{tab:table2}
  \end{center}
\end{table}

For the formation of Ps, we include radiative recombination with free
electrons and charge exchange with hydrogen and neutral He. For Ps
quenching, we include breakup of Ps by neutral hydrogen, ortho-Ps
quenching via free electrons and via electrons bound in H, inverse
charge exchange with H and He, and ionization of ortho-Ps atom by free
electrons. This last reaction turns out only to be important in
ionized gases at densities $\gtrsim 10^{13}$ cm$^{-3}$, and is
relevant to positron annihilation at the Sun.

For the Monte Carlo determination of the fraction of positrons forming
Ps before thermalization and the resulting Ps energy distribution, we
include He and we use an updated experimentally-determined cross
section for charge exchange in hydrogen. We find that this new cross
section results in higher Ps production than previously calculated,
and the resulting annihilation line width is broader.

The Ps formation, quenching and energy loss processes along with the
positron annihilation processes treated in our calculations are shown
in Table 2. The important diagnostic spectral features of annihilation
radiation are the ratio $Q_{3\gamma}/Q_{2\gamma}$ and the FWHM width
$\Delta E$ of the line. These two quantities depend upon the
conditions of the medium where the annihilation occurs. The medium can
be characterized by the temperature $T$, the total (neutral + ionized)
hydrogen density $n_H$, and the ionization fractions $X_{H^+}$,
$X_{He^{+}}$ and $X_{He^{++}}$. We use the non-thermal and
thermally-averaged rates discussed above to solve a system of
continuity equations for pair equilibrium and provide rates for
positron annihilation from all of the various processes. These rates
then directly give $Q_{3\gamma}/Q_{2\gamma}$. The total line spectrum
is the sum of several components, some having Gaussian shapes, others
not. Hence, the total line shape will not, in general, be Gaussian.

\begin{table}
  \begin{center}
    \caption{Temperatures and densities, in units of cm$^{-3}$,
 of different phases of the ISM used in calculations of 
annihilation radiation}\vspace{1em}
    \renewcommand{\arraystretch}{1.2}
    \begin{tabular}[h]{lcccc}
      \hline
       &  Cold & Warm & Warm  & Hot  \\
       &  Neutral & Neutral & Ionized  &   \\
      \hline
      T(K) & 80 & 8000 & 8000 & $4.5\times 10^5$  \\
	n$_{\rm HI}$ & 38 & 0.31 & 0.06 & 0.0 \\
      n$_e$ & 0.0 & 0.055 & 0.17 & $3.5\times 10^{-3}$  \\
      n$_{\rm HI}$ & 0.0 & 0.055 & 0.17 & $2.9\times 10^{-3}$ \\
      n$_{\rm HeI}$ & 3.8 & 0.036 & 0.023 & 0.0 \\
      n$_{\rm HeII}$ & 0.0 & 0.0 & 0.0 & 0.0 \\
      n$_{\rm HeIII}$ & 0.0 & 0.0 & 0.0 & $0.6\times 10^{-3}$ \\
      \hline \\
      \end{tabular}
    \label{tab:table3}
  \end{center}
\end{table}

\begin{figure}
\centering
\includegraphics[width=1.8\linewidth]{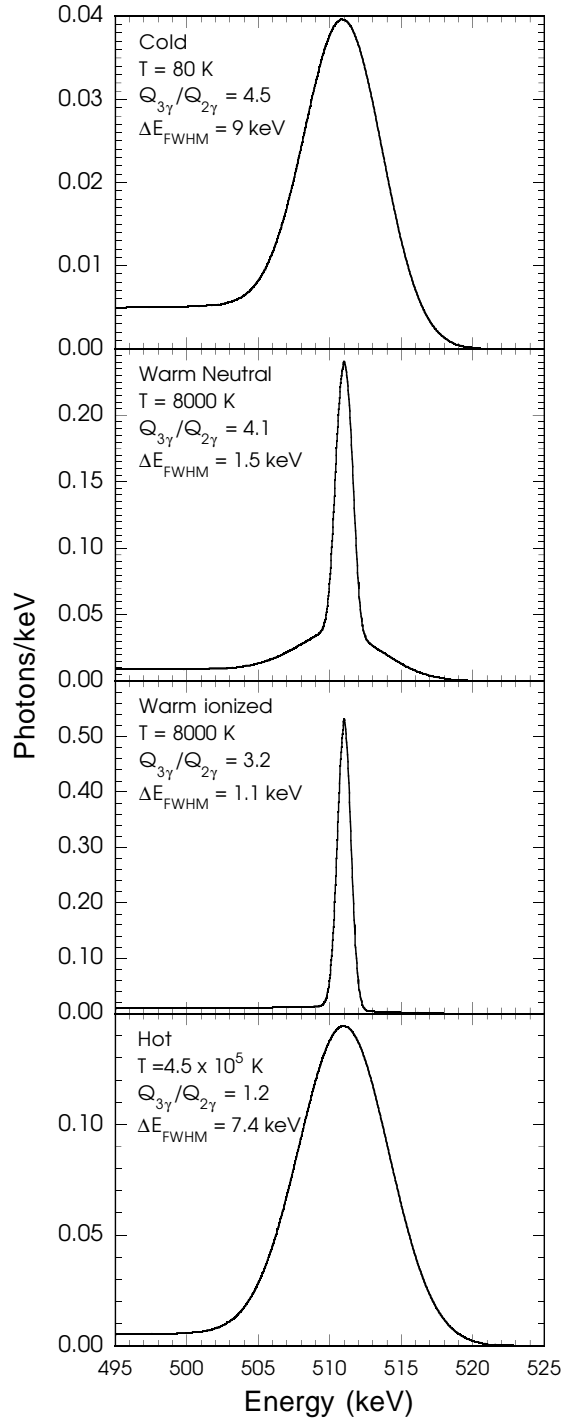}
\caption{Calculations of annihilation line and Ps continuum 
from the injection of MeV positrons into regions with different 
temperatures and ionization fractions.
\label{fig2}}
\end{figure}

Fig.\ 2 shows calculations of the annihilation line and Ps continuum
for phases of the ISM with temperatures and densities given by Table
3. Fig.\ 3 shows calculations of the same quantities for high
temperature, fully ionized media. The quantities
$Q_{3\gamma}/Q_{2\gamma}$, giving the relative number of photons in
the $3\gamma$ Ps continuum to the 0.511 MeV line, and the FWHM line
width $\Delta E$, from Figs.\ 2 and 3, are plotted in Fig. 4. In the
limit of a fully neutral medium, all annihilation takes place through
the formation of Ps. Consequently $Q_{3\gamma}/Q_{2\gamma}\rightarrow
9/2$, because 3/4 of the annihilations take place via ortho-Ps, giving
3 photons per annihilation, whereas 1/4 of the annihilations take
place via para-Ps, giving 2 photons per annihilation. Annihilation
occurs before thermalization in a fully neutral medium, so the
calculated 9 keV line width for the neutral medium is broader than for
partially ionized media where the positrons generally thermalize
before annihilating.

\begin{figure}
\centering
\includegraphics[width=1.0\linewidth]{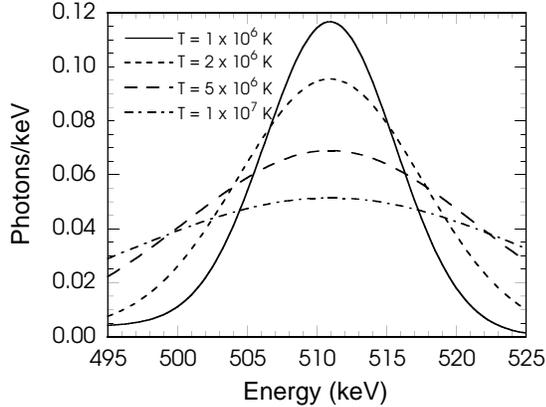}
\caption{Calculations of annihilation line and Ps continuum
 from the injection of MeV positrons into  high temperature plasma.
\label{fig3}}
\end{figure}

\begin{figure}
\centering
\includegraphics[width=1.0\linewidth]{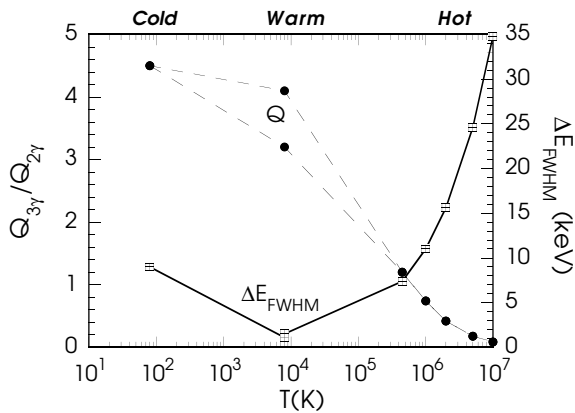}
\caption{Calculations of $3\gamma/2\gamma$ ratios 
and FWHM line widths in different phases of the ISM.
\label{fig4}}
\end{figure}

As the temperature increases, the relative number of $3\gamma$
continuum to 2$\gamma$ line photons decreases.  This is because the Ps
formation processes involving charge exchange with neutrals become
less important at higher temperatures due to the fewer available
neutral or partially ionized ions. The line width increases at high
temperatures due to thermal broadening, and approaches the limit
$\Delta E_{FWHM} = 11.0 T_6^{1/2}$ keV \citep{cra76,gsr97}, where
$T_6$ is the temperature in units of $10^6$ K.

Fig.\ 4 quantitatively shows how the $3\gamma$ Ps continuum is
quenched in high temperature plasmas. Consequently Ps formation can be
quenched in high temperature gas. This property can account for the
marginal detection or absence of a high-latitude enhancement in the Ps
maps of the Galactic Center region as compared with the observation of
a high-latitude enhancement in the 0.511 MeV line maps
\citep{mil99}. If this interpretation is correct, then a broadened
annihilation line would be detected in the direction of the
high-latitude enhancement with a line width $\Delta E_{FWHM} \gtrsim
10$ keV and with an intensity $\sim 1$-$2\times 10^{-4}$ ph cm$^{-2}$
s$^{-1}$. For broad FoV instruments such as TGRS \citep{har98}, this
broadened line is difficult to resolve from the narrow, bright
annihilation line emitted by the disk and bulge components. The
imaging capability of {\it INTEGRAL} will be able to determine whether
the interpretation of the high latitude enhancement as a fountain of
hot gas venting into the halo of the Galaxy is correct by mapping the
intensity and line width in this direction, provided that the positive
latitude enhancement is sufficiently localized. Further work will be
required to determine if Ps can be quenched by dust in the Galactic
Center region, which could also lead to suppression of the $3\gamma$
Ps continuum.

\section{Summary and Conclusions}

We have summarized our knowledge of astronomical sources of positrons
and the processes that lead to the formation of annihilation radiation
in the ISM. Because of the large uncertainties in nucleosynthesis
calculations, the relative contributions of novae, SNe, and Wolf-Rayet
winds remains unclear. Positrons from cosmic rays, compact objects,
and GRBs could also make a significant contribution to positron
production in the Galaxy. An important tool to reveal the dominant
sources of positrons will be to use $\gamma$-ray tracers to establish
associated positron injection rates. COMPTEL and {\it INTEGRAL} maps
of the $^{26}$Al 1.809 line and EGRET maps of $\pi^0$ gamma-ray
emission should trace some fraction of the annihilation glow of the
Galaxy, with the residual attributed to other processes.

One of the lingering uncertainties in positron astrophysics is the
question of whether positrons are formed by accreting compact objects
in the Galaxy. SPI on {\it INTEGRAL} will be well suited to examine
this question, not only by searching for annihilation line transients,
but also by looking for annihilation hot spots that are coincident
with compact object emission in the Galaxy. SPI's 511 keV sensitivity
for on-axis point sources during a $10^6$ s observation is 2-$3\times
10^{-5}$ ph cm$^{-2}$ s$^{-1}$ \citep{ved99}. Spectral analysis may
show that annihilation photons from directions coincident with known
compact objects have a different line width or $3\gamma/2\gamma$ ratio
than the surrounding emission, suggesting a point source origin. For
sufficiently bright sources, SPI could provide sub-degree imaging that
could be compared with IBIS maps of point sources. Searches for
variations of the Ps fraction and line width will reveal if there are
unusual sites of positron production, and whether the emission from
the positive latitude enhancement arises from annihilation in a hot
superwind or at compact objects.

\section*{Acknowledgments}

This work is supported by the Office of Naval Research. We thank
P. A. Milne and M. J. Harris for discussions.

% The following bibliography was produced with
%   \bibliographystyle{aa}
%   \bibliography{esapub}

\begin{thebibliography}{}

\bibitem[Bloemen et~al.(1984)]{blo84}
Bloemen, J.B.G.M., Blitz, L., Hermsen, W., 1984, ApJ, 279, 136

\bibitem[Bloemen et~al.(1999)]{blo99}
Bloemen, H., et~al., 1999, ApJ, 521, L137

\bibitem[Bussard et~al.(1979)]{bus79}
Bussard, R.W., Ramaty, R., Drachman, R.J., 1979, ApJ, 228, 928

\bibitem[Burdyuzha \& Kauts(1994)]{bur94}
Burdyuzha, V.V., Kauts, V.L., 1994, ApJS, 92, 549

\bibitem[Chan \& Lingenfelter(1993)]{cl93}
Chan, K.W., Lingenfelter, R.E., 1993, ApJ, 405, 614 

\bibitem[Cheng et~al.(1997)]{che97}
Cheng, L.X., et~al., 1997, ApJ, 481, L43

\bibitem[Crannell et~al.(1976)]{cra76}
Crannell, C.J., Joyce, G., Ramaty, R., Werntz, C., 1976, ApJ, 210, 582

\bibitem[Dermer \& Skibo(1997)]{ds97}
Dermer, C.D., Skibo, J.G., 1997, ApJ, 487, L57

\bibitem[Dermer \& Liang(1988)]{dl88}
Dermer, C.D., and Liang, E.P., 1988, in Nuclear Spectroscopy of
 Astrophysical Sources, ed. N. Gehrels and G.H. Share (New York: AIP), 326

\bibitem[Diehl et~al.(1995)]{die95}
Diehl, R., et~al., 1995, A\&A, 298, 445

\bibitem[Guessoum et~al.(1991)]{grl91}
Guessoum, N., Ramaty, R., Lingenfelter, R.E., 1991, ApJ, 378, 170

\bibitem[Guessoum et~al.(1997)]{gsr97}
Guessoum, N., Skibo, J.G., Ramaty, R., 1997, in the Second INTEGRAL 
Workshop, ed. C. Winkler, T. Courvoisier, Ph. Durouchoux (ESA SP-382), 113

\bibitem[Harris et~al.(1998)]{har98}
Harris, M.J., Teegarden, B.J., Cline, T.L., Gehrels, N., Palmer, D.M., 
Ramaty, R., Seifert, H., 
 1998, ApJ, 501, L55

\bibitem[for a review, see Harris(1997)]{har97}
Harris, M.J., 1997, in the Fourth Compton Symposium, ed. C.D. Dermer, 
M.S. Strickman, and J.D. Kurfess (New York: AIP), 418

\bibitem[Harris et~al.(1990)]{har90}
Harris, M.J., Share, G.H., Leising, M.D., Kinzer, R.L., Messina, D.C., 
1990, ApJ, 362, 135

\bibitem[Hernanz et~al.(1999)]{her99}
Hernanz, M., Jos\'e, J., Coc, A., G\'omez-Gomar, J., Isern, J., 1999,
 ApJ, 526, L97

\bibitem[Hernanz et~al.(1999a)]{her99a}
Hernanz, M., Jos\'e, J., Coc, A., G\'omez-Gomar, J., Isern, J., 1999,
 in the Fifth Compton Symposium, ed. M. L. McConnell and J. M. Ryan 
(New York: AIP), 97

\bibitem[Johnson \& Haymes(1973)]{jh73}
Johnson, W.N., Haymes, R.C., 1973, ApJ, 184, 103

\bibitem[Kinzer et~al.(1999)]{kin99}
Kinzer, R.L, Purcell, W.R., Kurfess, J.D., 1999, ApJ, 515, 215

\bibitem[Leventhal et~al.(1978)]{lev78}
Leventhal, M., MacCallum, C.J., Stang, P.D., 1978, ApJ, 225, L11 

\bibitem[Li \& Liang(1996)]{ll96}
Li, H., Liang, E.P., 1996, ApJ, 458, 514

\bibitem[Lingenfelter \& Ramaty(1989)]{lr89}
Lingenfelter, R.E., Ramaty, R., 1989, ApJ, 343, 686

\bibitem[Mahoney et~al.(1994)]{mlw94}
Mahoney, W.A., Ling, J.C., Wheaton, W.A., 1994, ApJS, 92, 387

\bibitem[Meynet et~al.(1997)]{mey97}
Meynet, G., Arnould, M., Prantzos, N., Paulus, G., 1997, A\&A, 320, 460

\bibitem[Milne et~al.(1999)]{mil99}
Milne, P.A., Kurfess, J.D., Kinzer, R.L., Leising, M.D., Dixon, D.D., 
1999, in the Fifth Compton Symposium, ed. M.L. McConnell and J.M. Ryan 
(New York: AIP), 21

\bibitem[Milne et~al.(1999a)]{mtl99}
Milne, P.A., The, L.S., Leising, M.D., ApJS, 124, 504

\bibitem[Phlips et~al.(1996)]{phl96}
Phlips, B.F., et~al., ApJ, 465, 907

\bibitem[Prantzos(1996)]{pra96}
Prantzos, N. 1996, A\&AS, 120, 303

\bibitem[Purcell et~al.(1997)]{pur97}
Purcell, W.R., et~al., ApJ, 491, 725

\bibitem[Ramaty et~al.(1994)]{rsl94}
Ramaty, R., Skibo, J.G., Lingenfelter, R.E., 1994, ApJS, 92, 393

\bibitem[Skibo(1993)]{ski93}
Skibo, J.G., 1993, PhD Thesis, Univ. of Maryland

\bibitem[Share et~al.(1990)]{sha90}
Share, G.H., Leising, M.D., Messina, D.C., Purcell, W.R., 1990, ApJ, 358, L45

\bibitem[Share et~al.(1988)]{sha88}
Share, G.H., et~al., 1988, ApJ, 326, 717

\bibitem[Starrfield et~al.(1997)]{sta97}
Starrfield, S., Truran, J.W., Wiescher, M.C., Sparks, 
W.M., 1997, in the Fourth Compton Symposium, ed. C.D.
 Dermer, M.S. Strickman, and J.D. Kurfess (New York: AIP), 1130

\bibitem[Vedrenne et~al.(1999)]{ved99}
Vedrenne, G., et~al.,  1999, Astron. Lett. and Communications, 39, 325

\bibitem[Wallyn et~al.(1996)]{wal96}
Wallyn, P., Mahoney, W.A., Durouchoux, P., Chapuis, C., 1996, ApJ, 465, 473

\bibitem[Woosley \& Weaver(1992)]{ww92}
Woosley, S.E., Weaver, T.A. 1992, in Supernovae, ed. J. Audouze et~al. (New York: Elsevier)

\bibitem[Zurek(1985)]{zur85}
Zurek, W.H. 1985, ApJ, 289, 603

\end{thebibliography}
% The results are inserted directly here to simplify
% the demonstration.

\end{document}